\documentclass[aps,prl,floatfix,superscriptaddress,showpacs,twocolumn]{revtex4}

\usepackage{longtable}
\usepackage{bbm}
\usepackage{times}
\usepackage{nicefrac}
\usepackage{amsfonts}
\usepackage[ansinew]{inputenc}
\usepackage{epsfig}
\usepackage{graphicx}
\usepackage{adjustbox}
\usepackage{color}
\usepackage{bm}
\usepackage{multirow}
\usepackage{mathrsfs}
\usepackage{revsymb}
\usepackage{amssymb}
\usepackage{amsmath}
\usepackage{mathtools}
\usepackage{slashed}
\usepackage{longtable}
\usepackage{dcolumn}
\usepackage[babel]{csquotes}

\newcolumntype{.}{D{x}{}{-1}}

\begin{document}

\title{Improved bound-electron \textit{g}-factor theory through complete two-loop QED calculations}

\author{B. Sikora}
\email[]{bastian.sikora@mpi-hd.mpg.de}
\author{V.~A. Yerokhin}
\author{C.~H. Keitel}
\author{Z. Harman}
\email[]{harman@mpi-hd.mpg.de}

\address{Max~Planck~Institute for Nuclear Physics, Saupfercheckweg~1, D~69117 Heidelberg, Germany}

\begin{abstract}

The two-loop self-energy correction to the bound-electron $g$-factor in hydrogenlike ions is investigated, taking into account the electron-nucleus interaction exactly. This all-order calculation is required to improve the total theoretical uncertainty of the $g$-factor, which is limited by the fact that two-loop self-energy corrections have only been calculated so far in the form of an expansion in $Z\alpha$. 
Here, $Z$ is the nuclear charge number and $\alpha$ is the fine-structure constant. 
In this work, we report calculations of the last missing parts of the total two-loop self-energy correction, exactly in $Z\alpha$. We apply our theory to the recently measured $g$-factor of the hydrogenlike $^{118}$Sn$^{49+}$ ion~[J. Morgner \textit{et al.}, Nature \textbf{622}, 53 (2023)] and, with a factor of 8, improve the accuracy of its state-of-the-art theoretical value by almost one order of magnitude, enabling more detailed tests of quantum electrodynamics and new physics in strong fields.

\end{abstract}


\maketitle

The $g$-factor of the electron has been an outstanding testing ground for the theory of quantum electrodynamics (QED)~\cite{Gabrielse2023}. Theory calculations of the free electron's $g$-factor involve Feynman diagrams with up to 5  loops~\cite{Laporta1996,Laporta2017,Kinoshita2012,Volkov2019}. 
Measurements of the bound electron's $g$-factor have allowed precision tests of the theory of QED in the presence of strong electric (nuclear) background fields~\cite{Karshenboim2000,Czarnecki00,Pachucki05,Czarnecki2017,Yerokhin04,Karshenboim2001,Beier00,Sturm11,Sturm13,Koehler2016,Arapoglou2018}  and were used to improve the precision of the electron mass~\cite{Sturm14,Koehler2015,Zatorski2017}. Furthermore, an improved determination of the fine-structure constant might be possible in the future~\cite{Shabaev2006,Yerokhin2016,Cakir2020}, as well as new physics searches~\cite{Debierre2020,Sailer2022}. 

Recent experiments tested the bound-electron $g$-factor theory in 
hydrogenlike $^3$He~\cite{Schneider2022}, $^9$Be~\cite{Dickopf2024}, $^{20,22}$Ne~\cite{Heisse2023} and $^{118}$Sn~\cite{Morgner2023}, the latter being the heaviest element for which a $g$-factor measurement has been performed with high precision. 
While the experimental and theoretical $g$-factor values for hydrogenlike $^{118}$Sn$^{49+}$ were found to be in agreement, the theoretical uncertainty in that work was much larger than the experimental one, and the theoretical uncertainty is dominated by uncalculated two-loop QED corrections of $\mathcal{O} \left( (Z\alpha)^6 \right)$. 
These uncalculated two-loop QED corrections determine the overall theoretical uncertainty already for $Z=6$~\cite{Czarnecki2017}, and are significantly larger than the experimental uncertainty for $Z=14$~\cite{Sturm13,Czarnecki2017}. Experimental efforts are underway to measure $g$-factors in ions heavier than Sn~\cite{Sturm2019,Herfurth2015}. In these systems, the uncertainty due to uncalculated two-loop QED terms is expected to strongly increase~$\propto Z^6$, making a precise theory prediction nearly impossible.

\begin{figure}[ht]
\begin{center}
\begin{tabular}{llclclc}
N:  & $a$ & \adjincludegraphics[valign=c,width=0.12\textwidth]{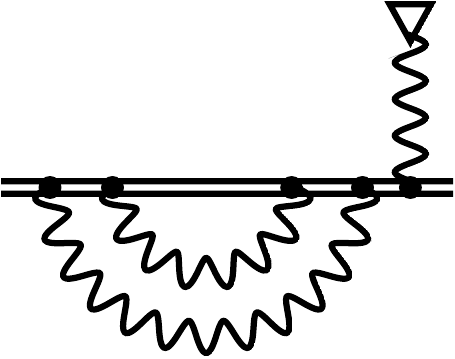}           & $b$ &
       \adjincludegraphics[valign=c,width=0.12\textwidth]{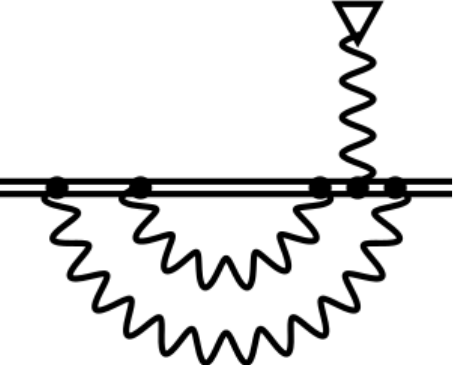}  & $c$ &
       \adjincludegraphics[valign=c,width=0.12\textwidth]{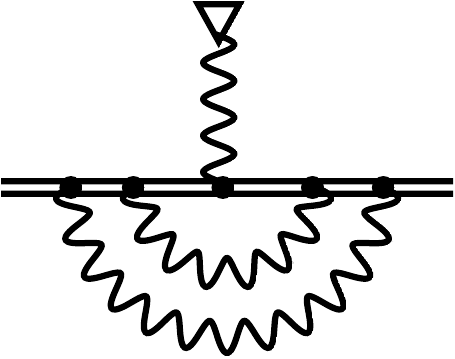} \\[3mm]
O:   & $d$ & \adjincludegraphics[valign=c,width=0.12\textwidth]{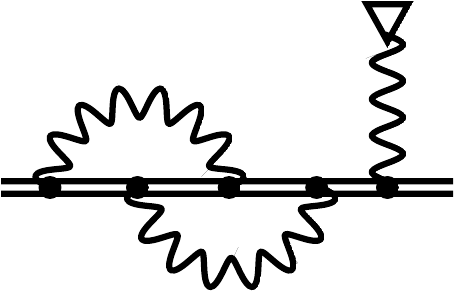}           & $e$ &
       \adjincludegraphics[valign=c,width=0.12\textwidth]{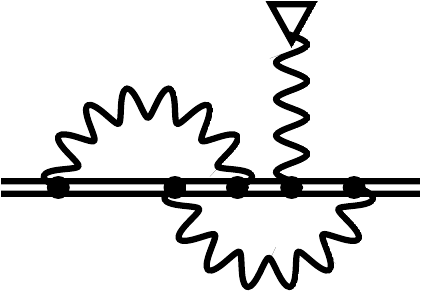} &     $f$ &
       \adjincludegraphics[valign=c,width=0.10\textwidth]{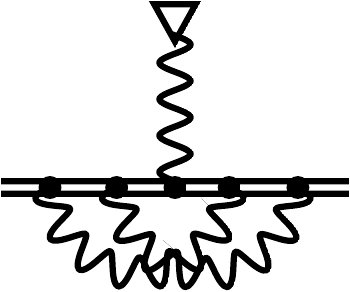}
\end{tabular}
\end{center}
\caption{Nested loop (N) and overlapping loop (O) Feynman diagrams of the two-loop self-energy corrections to the bound-electron $g$-factor \cite{Beier00}. Double lines represent the bound electron; the wave line denotes the virtual photon. 
A wave line terminated by a triangle represents a magnetic interaction.}
\label{fig:gSESE}
\end{figure}

This necessitates large-scale theory efforts to calculate QED Feynman diagrams with two loops, taking into account the electron-nucleus interaction exactly, i.e. without expansion in $Z\alpha$. There were considerable efforts invested in this project~\cite{Yerokhin2Loop2013,Debierre2021}. However, the dominant part of these effects, Feynman diagrams with two self-energy loops (SESE diagrams), shown in Fig.~\ref{fig:gSESE}, remained uncalculated so far and have represented an open challenge which is solved in this letter. 

Self-energy corrections are typically ultraviolet (UV) divergent, thus these divergenes have to be carefully isolated. Renormalization methods have
been elaborated in momentum space for diagrams containing
free Dirac propagators, while the Dirac Coulomb Green's functions (DCGF)
are only known in coordinate space. Therefore, we separate contributions with the DCGF replaced by propagators containing zero or one
interaction with the Coulomb field in such a way that the corresponding difference is rendered UV finite. 
In addition to the loop-after-loop (LAL) diagrams, in which one self-energy loop closes before the second one opens, two-loop diagrams can to be
cast into three different categories: (i) terms which contain UV divergences, (ii) terms which contain the DCGF, and (iii) terms which contain
both DCGF and UV divergences through a one-loop subgraph. Using the nomenclature from Ref.~\cite{Mallampalli1998}, we refer to these categories as the F, M, and P terms, respectively, which all require different analytical and numerical methods.

\textit{Nested and overlapping loops} -- The nested (N) and overlapping (O) loop Feynman diagrams are shown in Fig.~\ref{fig:gSESE}. These diagrams represent the main calculational challenge. First, there are the wave function corrections [Fig.~1~(a) and (d)] where one of the external electron lines is perturbed by the magnetic interaction. 
 The electron propagator between the magnetic interaction and the SE loops can be represented as a sum over the spectrum of the Coulomb-Dirac Hamiltonian, schematically
$G(E ) = \sum \limits_n \frac{|n \rangle \langle n |}{E - E_n}$, with the $E_n$ being eigenenergies of the eigenstates $|n\rangle$. The cases $E_n = E_{1s}$ and $E_n \neq E_{1s}$ need to
be treated separately. Following the usual convention, we call these two contributions the reducible (\enquote{red}) and the irreducible (\enquote{irred}) parts, respectively, since they can or cannot be reduced to expressions originating from a lower order of perturbation theory.
The corresponding energy shift of the irreducible corrections can be represented as
\begin{align}
\Delta E _{i, \rm wf,irred} = 2 \langle \delta \psi \vert \gamma^0 \Sigma_i \vert \psi \rangle  \, ,
\label{eq:gwfirred}
\end{align}
where $i\in\{N,O\}$ stands for nested and overlapping loop contributions respectively. $\vert \delta \psi \rangle$ represents the electron's wave function perturbed by the external magnetic field~\cite{ShabaevVirial} and $\gamma^0$ is a Dirac matrix in the standard representation. The corresponding two-loop self-energy functions $\Sigma_i$ were previously derived for the calculation of the two-loop SE correction to the Lamb shift (e.g.~\cite{Yerokhin2003}) and independently in our previous work~\cite{Sikora2020}. 

The $g$-factor contribution of the reducible part of the wave function corrections can be written as
\begin{align}
g _{i, \rm red} = g_D \langle  \psi \vert \gamma^0 \left. \frac{\partial \Sigma_i}{\partial E} \right|_{E = E_{1s}} \vert \psi \rangle  \, ,
\end{align}
where $g_D$ is the Dirac value of the $g$-factor~\cite{Breit1928}. Depending on whether the derivative acts on the central electron propagators or one of the \enquote{side} electron propagators, we label the reducible contribution as \enquote{ladder} or \enquote{side}, following the nomenclature of Ref.~\cite{Yerokhin2003}.

The energy shift corresponding to vertex diagrams can be schematically represented as
\begin{align}
\Delta E _{\mathrm{ver},i,j} = \langle \psi \vert \gamma^0 \boldsymbol{\Gamma}_{i,j} \boldsymbol{A} \vert \psi \rangle  \, ,
\end{align}
with the two-loop vertex functions $\boldsymbol{\Gamma}_{i,j}$ derived in our previous work~\cite{Sikora2020,SikoraThesis} and the vector potential $\boldsymbol{A}$ corresponding to a constant external magnetic field. $j$ stands for either \enquote{side} or \enquote{ladder}. 

Every N and O diagram was split into F-, M- and P-term parts. The scheme according to which full CDGF need to be replaced by $n$-potential Green functions to obtain the corresponding F-, M- or P-term part of all diagrams is schematically tabulated in the Supplement.

\textit{F-term} -- The F-term consists of the UV divergent parts of N and O diagrams. 
The irreducible wave function corrections consist of zero- and one-potential contributions, all other diagrams only have zero-potential contributions to the F-term. 
%
%
%
%
%
Once the zero-potential SE functions for the computation of the irreducible wave function corrections were calculated, it was straightforward to calculate their derivatives with respect to energy, necessary for the computation of the reducible parts.

The computation of vertex diagrams was performed using the magnetic potential in momentum space~\cite{Yerokhin04}
%
$\boldsymbol{A} (\boldsymbol{p}_1 - \boldsymbol{p}_2) = - \frac{i}{2} (2 \pi)^3 \boldsymbol{B} \times \nabla_{p_2} \delta (\boldsymbol{p}_1 - \boldsymbol{p}_2) \, .$
%
To resolve the derivative of the delta function, we perform an integration by parts. This leads to two types of vertex diagram contributions in which the derivative acts on the vertex function and on the electron wave function, respectively. 
For the calculation of the part with the derivative acting on the wave functions, we use the Ward identity ~\cite{QFT}
\begin{align}
\sum_j {\Gamma_{ij}^{(0)}}^\mu (p,p) = & - \partial^\mu \Sigma_i^{(0)} (p) 
\end{align}
which is a generalization of the one-loop case~\cite{Yerokhin04}. This means that also this part of vertex contributions can be calculated straightforwardly from the zero-potential SE functions. 
The derivatives of the vertex functions 
\begin{align}
\boldsymbol{\Xi}_{ij} (p) = \left. \nabla_{p_2} \times \boldsymbol{\Gamma}^{(0)}_{ij \mathrm{R}} (p_1, p_2)_z \right|_{p_2=p_1} \, 
\end{align}
were calculated in two different ways. First, we derived momentum integral formulas for these entities and solved these integrals using dimensional regularization and the Feynman parameter technique. 
As a consistency test, we calculated the zero-potential contributions to the vertex functions using Feynman parameters and calculated the necessary derivatives from them. Schematic formulas are presented in the supplement. 
Results from both approaches were consistent with one another.

\textit{M-term} -- 
The M-term contributions of N and O diagrams are the UV finite parts of these diagrams. The main challenges in M-term calculation are described below, namely, the proper handling of all infrared (IR) divergences, a double infinite summation over angular momentum quantum numbers as well as a multidimensional integration which needs to be performed numerically. 


IR divergences occur in vertex and reducible wave function contributions in the case of (at least) two electron propagators inside SE loops which correspond to the same energy. This occurence of IR divergences is well known for the case of the one-loop SE correction to the $g$-factor~\cite{Yerokhin04} and the two-loop SE correction to the Lamb shift~\cite{Yerokhin2003}. In case of the one-loop SE correction to the $g$-factor, IR divergences are dealt with by directly calculating the sum of two IR divergent contributions whose IR divergences cancel one another. We were able to use this approach in the calculation of the M-term contributions to O diagrams. Specifically, both in the case of the O, vertex, side and the O, vertex, ladder diagrams, the IR divergences are cancelled by directly calculating their sum with the corresponding O, red contributions (referring to such a combination as \enquote{VR} in the following).


For the N, vertex diagrams, calculating directly the sum of the vertex diagram and the corresponding reducible wave function correction softens the IR divergence, but some IR divergences remain for which we apply the method developed for the SESE correction to the Lamb shift~\cite{Yerokhin2018twoloop}. 
For numerical calculations, the remaining IR divergent terms were subtracted from both the N, VR contributions, and numerical values of the finite remainders are computed explicitly. 



\textit{Numerical aspects} -- Every M-term contribution of an N and O diagram contains a double infinite summation over angular momentum quantum numbers. 
Inside vertex diagrams, there are four electron propagators, each represented by a partial-wave expansion with the angular-momentum parameter $\kappa$. The application of angular-momentum selection rules leaves two of the $\kappa$s unbounded, whereas for the other two we obtain the following restrictions:
\begin{itemize}
\item for N, vertex, side: $\kappa_3=\kappa_1, \kappa_4 \in \{ \kappa_1, -\kappa_1-1,-\kappa_1+1\}\, ,$
\item for N, vertex, ladder: $\kappa_3\in \{ \kappa_2, -\kappa_2-1,-\kappa_2+1\}, \kappa_4 \in \{ \kappa_1, -\kappa_1-1,-\kappa_1+1\} \, ,$
\item for O, vertex, side: $\kappa_4 \in \{ \kappa_3, -\kappa_3-1,-\kappa_3+1\}$, $\vert \kappa_2 \vert \in \{ \vert \kappa_1 - \kappa_3 \vert , \cdots , |\kappa_1|+|\kappa_3| \} \, ,$
\item for O, vertex, ladder: $\kappa_3 \in \{ \kappa_2, -\kappa_2-1,-\kappa_2+1\}$, $\vert \kappa_2 \vert \in \{ \vert \kappa_1 - \kappa_4 \vert , \cdots , |\kappa_1|+|\kappa_4| \}\, .$
\end{itemize}
Note that in O diagrams, the sum over $\kappa_2$, while finite, involves an increasingly large number of terms when $\kappa_1$ and $\kappa_4$ get large, increasing the numerical computation time accordingly. Partial waves for O diagrams were calculated for all $l_1 = \left| |\kappa_1| - |\kappa_4| \right| + 1 \in \{ 1,\ldots , 8 \}$ and $l_2 = \min (|\kappa_1|,|\kappa_4|) \in \{ 1, \ldots , 7 \}$, more partial waves were computed for N diagrams.  The limiting factor on the achievable precicion of the M-term is the convergence of the partial wave expansion of the O, vertex diagrams. The extrapolation procedure for the double infinite summation was described in Ref.~\cite{Yerokhin2018twoloop}. 

On top of the double infinite summation, every M-term contribution requires a multidimensional integration to be performed numerically, making the calculation of the M-term the computationally most demanding part of the total SESE calculation. After performing the angular integrations of all radial variables analytically, we are left with two frequency integrations (one for each SE loop) and up to five radial integrations (one for each vertex) to be performed numerically, for each term of the partial-wave expansion. 


\textit{P-term} -- The P term is represented by Feynman diagrams containing both the
ultraviolet (UV) divergent subgraphs {\em and} the bound-electron propagators.
Such diagrams are a computational challenge because 
they need to be computed in the mixed momentum-coordinate representation, with
a part of each diagram evaluated in momentum space and the 
remaining part, in coordinate space. 

After the UV renormalization, the individual P-term diagrams still diverge, 
due to inherent 
infrared (IR) singularities. These divergences
are regularized by introducing the finite photon mass $\mu$. The divergencies
are identified and isolated by the method described in 
Ref.~\cite{yerokhin:20:green} and parameterized in terms of $\ln \mu$ and $1/\mu$.
The cancelation of the $\mu$-dependent terms is carried out analytically,
leaving finite expressions for each diagram to be computed numerically.

For the numerical computation of the P term we generalize the numerical method developed in 
Ref.~\cite{Yerokhin2010} based on 
the analytical representation of the DCGF in terms 
of the Whittaker functions. 
The crucial part of the method is
the Fourier transform of DCGF
over one of the radial arguments. We start with the known representation
of the radial DCGF in terms
of the two-component solutions of the radial Dirac equation
which are regular at the origin $\left(\phi_{\kappa}^{0}\right)$
and the infinity $\left(\phi_{\kappa}^{\infty}\right)$ 
\cite{mohr:00:rmp},
\begin{align}\label{gr01}
 G_{\kappa}(E,x_1,x_2) = &\,
 -\phi_{\kappa}^{\infty}(E,x_1)\,\phi_{\kappa}^{0^T}(E,x_2)\,\theta(x_1-x_2)
\nonumber \\ &
 -\phi_{\kappa}^{0}(E,x_1)\,\phi_{\kappa}^{{\infty}^T}(E,x_2)\,\theta(x_2-x_1)\,,
\end{align}
where $\theta(x)$ is the Heaviside step function and $T$ denotes the transpose, $x_1$ and $x_2$ are the radial coordinates. 
Now, the Fourier transform of $G_{\kappa}(E,x_1,x_2)$ over, e.g., the second radial
argument is expressed as
\begin{align}\label{gr2}
 G_{\kappa}(E,x_1,p_2)  = &\,
 -\phi_{\kappa}^{\infty}(E,x_1)\,\psi_{\kappa}^{0^T}(E,x_1,p_2)
\nonumber \\ &
 -\phi_{\kappa}^{0}(E,x_1)\,\psi_{\kappa}^{{\infty}^T}(E,x_1,p_2)\,,
\end{align}
\begin{align}\label{gr3}
\psi_{\kappa}^{0}(E,x_1,p_2) = &\, 4\pi\,
\int_0^{x_1}\!\!\!\! dx_2\,
  x_2^2\,
\left(
        \begin{array}{r}
   j_l(p_2x_2)\,\phi^0_{\kappa,+}(E,x_2) \\[0.5em]
   a_{\kappa}\,j_{\overline{l}}(p_2x_2)\,\phi^0_{\kappa,-}(E,x_2) \\
        \end{array}
\right)\,,
\end{align}
\begin{align}\label{gr4}
\psi_{\kappa}^{\infty}(E,x_1,p_2) = &\, 4\pi\,
\int_{x_1}^{\infty}\!\!\!\! dx_2\,
  x_2^2\,
\left(
        \begin{array}{r}
   j_l(p_2x_2)\,\phi^{\infty}_{\kappa,+}(E,x_2) \\[0.5em]
   a_{\kappa}\,j_{\overline{l}}(p_2x_2)\,\phi^{\infty}_{\kappa,-}(E,x_2) \\
        \end{array}
\right)\,,
\end{align}
where $a_{\kappa} = -{\rm sign}(\kappa)$, $\phi_{\kappa,+}$ and $\phi_{\kappa,-}$ are
the upper and smaller components of the functions $\phi_{\kappa}$, respectively,
$l = |\kappa+1/2|-1/2$,  $\overline{l} = |\kappa-1/2|-1/2$, and $j_l(z)$ are the
spherical Bessel functions. 

The functions $\psi_{\kappa}^0$ and
$\psi_{\kappa}^{\infty}$ are the most computationally intensive parts 
of the calculations since the integration 
over $x_2$ involves spherical Bessel functions which
oscillates rapidly for large values of $p_2$. For our computations, we had
to develop a procedure for the numerical Fourier transform with
a controllable accuracy for momenta as high as $p=10^6$.
The crucial part of the computational scheme was to avoid multiple evaluations of 
the Fourier-transform integrals with different values of $x_1$. This was
achieved by pre-storing an ordered radial grid $\left\{x_{1,i}\right\}$
and computing the whole set 
of values $\left\{\psi_{\kappa}(E,x_{1,i},p_2)\right\}$ for a given $p_2$
by performing just one Bessel transform over the interval $x_1 \in(0,\infty)$. 

\begin{figure}
\begin{center}
\begin{tabular}{crcccrcccc}
($a$) & $2\times$ & \adjincludegraphics[valign=c,width=0.08\textwidth]{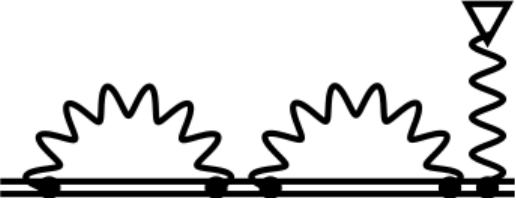} & \quad\quad &
($b$) & $2\times$ & \adjincludegraphics[valign=c,width=0.08\textwidth]{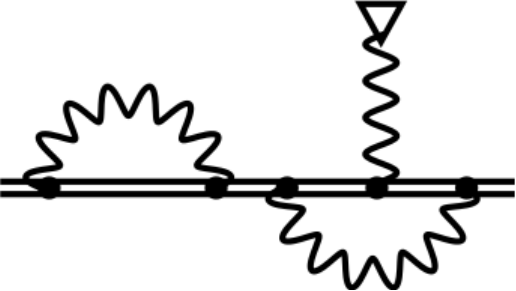}   & \quad\quad &
($c$) &             \adjincludegraphics[valign=c,width=0.08\textwidth]{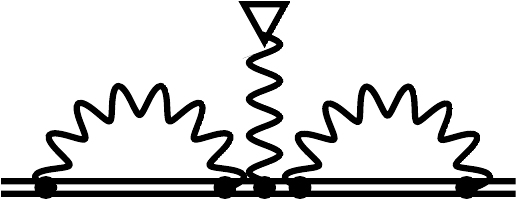}  \\
 & & & & & & & & &
\end{tabular}
\end{center}
\begin{center}
\begin{tabular}{crcccrcccc}
($d$) & $2\times$ & \adjincludegraphics[valign=b,width=0.04\textwidth]{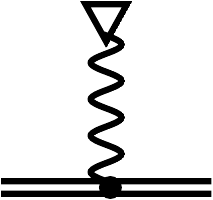}                & $\times$   &
                  \adjincludegraphics[valign=b,width=0.08\textwidth]{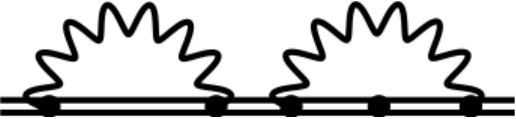} & \quad\quad &
($e$) &             \adjincludegraphics[valign=b,width=0.04\textwidth]{g-factor}                & $\times$   &
                  \adjincludegraphics[valign=b,width=0.08\textwidth]{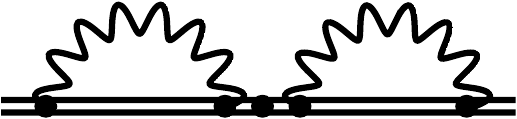}
\end{tabular}
\end{center}
\caption{LAL contribution of the SESE correction. Double lines between two SE loops or between an SE loop and the magnetic interaction represent reduced Green's functions. A double line with a dot denotes a derivative
of a (full or reduced) Green's function with respect to the energy: $\left. \frac{\partial G (E)}{\partial E} \right|_{E=E_{1s}}$. }
\label{fig:LALdiagrams}
\end{figure}

\textit{LAL} -- Following our analysis~\cite{SikoraThesis,Sikora2020} using the two-time Green's function method~\cite{Shabaev2002Report}, we group all contributions from Feynman diagrams in which one SE loop closes before the second SE loop opens into the \enquote{loop after loop, irreducible} and the \enquote{loop after loop, reducible} (LAL, red) part of the total SESE correction. 
It is the \enquote{loop after loop, irreducible} contribution which we will refer to as LAL. The different LAL contributions are shown in Fig.~\ref{fig:LALdiagrams}. We calculated these diagrams with the help of bound-electron wave functions perturbed by a self-energy loop, $\vert \Psi_{\Sigma} \rangle$~\cite{Holmberg15}. Using this self-energy perturbed wave function (SEWF), all LAL, irreducible contributions can be calculated as straightforward generalizations of the one-loop SE calculations, see Refs.~\cite{Yerokhin04,Sikora2020,SikoraThesis} for details. 

\begin{figure}
\begin{center}
\begin{tabular}{cccccccccccccc}
($a$) & \adjincludegraphics[valign=b,width=0.06\textwidth]{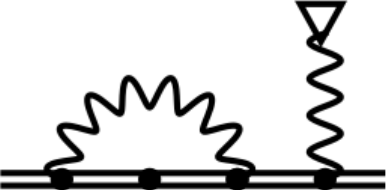}     & $\times$   &
      \adjincludegraphics[valign=b,width=0.04\textwidth]{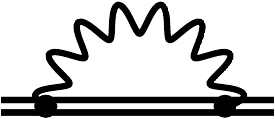}                       &  &
($b$) & \adjincludegraphics[valign=b,width=0.06\textwidth]{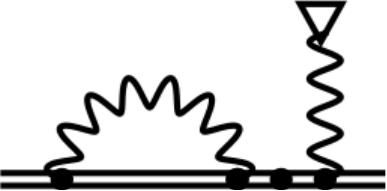} & $\times$   &
      \adjincludegraphics[valign=b,width=0.04\textwidth]{Lamb-SE}                       &  &
($c$) & \adjincludegraphics[valign=b,width=0.06\textwidth]{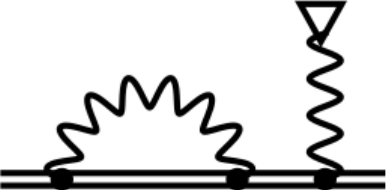}              & $\times$
      \adjincludegraphics[valign=b,width=0.04\textwidth]{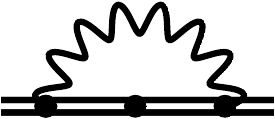}                   \\
 & & & & & & & & & & & & &
\end{tabular}
\end{center}
\begin{center}
\begin{tabular}{ccccccccc}
($d$) & \adjincludegraphics[valign=c,width=0.05\textwidth]{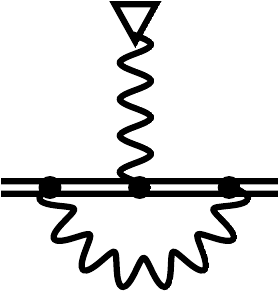}          & $\times$   &
      \adjincludegraphics[valign=c,width=0.05\textwidth]{Lamb-SE-dot}                   & \quad\quad &
($e$) & \adjincludegraphics[valign=c,width=0.05\textwidth]{Lamb-SE}                     & $\times$   &
      \adjincludegraphics[valign=c,width=0.05\textwidth]{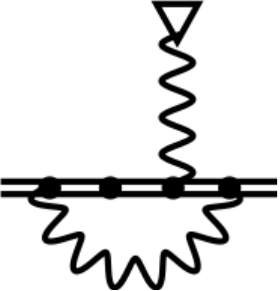}        \\
 & & & & & & & &
\end{tabular}
\begin{tabular}{ccccccccccccc}
($f$) & \adjincludegraphics[valign=b,width=0.04\textwidth]{g-factor}                    & $\times$   &
      \adjincludegraphics[valign=b,width=0.05\textwidth]{Lamb-SE-dot}                   & $\times$   &
      \adjincludegraphics[valign=b,width=0.05\textwidth]{Lamb-SE-dot}                   & \quad\quad &
($g$) & \adjincludegraphics[valign=b,width=0.04\textwidth]{g-factor}                    & $\times$   &
      \adjincludegraphics[valign=b,width=0.05\textwidth]{Lamb-SE}                       & $\times$   &
      \adjincludegraphics[valign=b,width=0.05\textwidth]{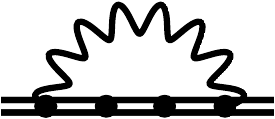}
\end{tabular}
\end{center}
\caption{Reducible two-loop SE diagrams which can be represented as products of one-loop diagrams (\enquote{LAL, red} contribution).}
\label{fig:LALreddiagrams}
\end{figure}

\textit{LAL, reducible} -- 
We refer to contributions which can be described as the product of two one-loop Feynman diagrams as the LAL, reducible (\enquote{LAL, red}) term, shown in Fig.~\ref{fig:LALreddiagrams}. We calculated these contributions by splitting one-loop diagrams into zero-, one- (where necessary) and many-potential terms. These include the derivative of the one-loop SE matrix element and the derivative of the one-loop SE correction to the $g$-factor. We compared our results with established literature values wherever possible. 
Specifically, we compare our results to the one- and two-loop self-energy corrections to the Lamb shift~\cite{Yerokhin2005,Yerokhin2018twoloop} and the one-loop self-energy correction to the $g$-factor~\cite{Yerokhin04}. 
Infrared (IR) divergences occur in some LAL, red contributions in the case of two electron propagators inside the SE loop which correspond to the same energy. In such cases, we introduced subtraction terms to cancel the IR divergence. 
Numerical results are then calculated for the finite remainder of the LAL, red contribution. The subtracted IR divergent terms were (analytically) combined with IR divergent terms from the N diagrams. The sum of IR terms of N and LAL, red contributions is finite and then calculated numerically. A detailed list of IR divergent terms from LAL, red and N (with mutual cancellations pointed out) is presented in the supplement. 

\begin{figure}
\begin{center}
\includegraphics[width=0.4\textwidth]{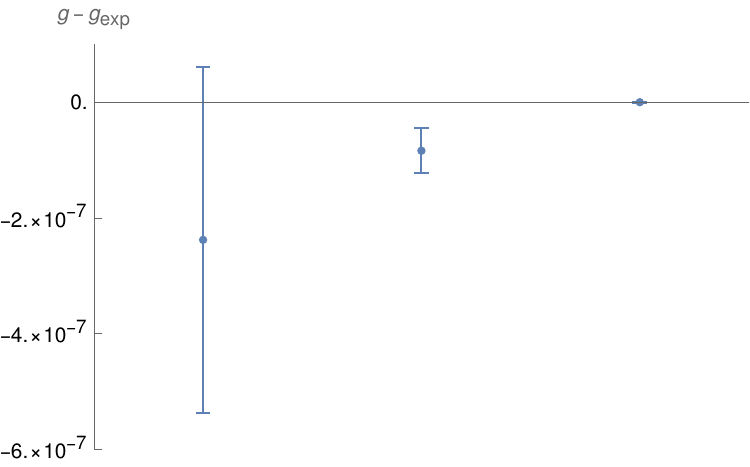}\\
$\quad$ theo, prev $\qquad \quad$  theo,new $\quad \qquad $ exp 
\end{center}
\caption{Previous and updated theoretical values of the bound-electron $g$-factor in $^{118}$Sn$^{49+}$, compared with the experimental $g$-factor value from Ref.~\cite{Morgner2023}.}
\label{fig:SESEplot}
\end{figure}

\textit{Results} -- In the present work, we apply our completed theory of the two-loop self-energy correction to hydrogenlike Sn. Table~\ref{tab:SESE} lists the results for the individual contributions to the SESE correction. In total, we find a SESE correction to all orders in $Z\alpha$ of $-4.099(19)\times 10^{-6}$. The major part of the uncertainty stems from the M-term contribution of O diagrams. We compare our results with the previously available theory of the SESE correction based on $Z\alpha$ expansion, summarized in Table~\ref{tab:SESENRQED}. 
 Subtracting the known $(Z\alpha)$ expansion results from our all-order result, we obtain the following higher-order SESE term $g_{\rm SESE}^{(Z\alpha)^{6+}}=1.53(19)\times 10^{-7}$. The previously published $g$-factor of H-like Sn is~\cite{Morgner2023} 
\begin{align}
g_{\rm Sn,theo,prev} = 1.910 \, 561 \, 821 (299) \, ,
\end{align}
with the estimated value and uncertainty due to the higher-order SESE term being $g_{\rm SESE,prev}^{(Z\alpha)^{6+}}=0 \pm 2.97 \times 10^{-7}$. With our calculation, we can now update the theoretical $g$-factor value of the H-like $^{118}$Sn$^{49+}$:
\begin{align}
g_{\rm Sn,theo} = 1.910 \, 561 \, 975  (39)\, .
\end{align}
The updated theoretical uncertainty is almost one order of magnitude smaller than the one published in Ref.~\cite{Morgner2023}.  The numerical uncertainty of the updated theoretical value comes from several sources: (i)  the SESE correction ($1.9 \times 10^{-8}$), (ii) the nuclear effects ($2.4 \times 10^-8$), (iii) the uncalculated two-loop effects with magnetic-loop vacuum polarization ($2.2 \times 10^-8$) (iv) uncalculated three-loop binding corrections ($1.1 \times 10^-8$)~\cite{Morgner2023}. 
The updated theoretical value and the experimental value agree within $2.1\sigma$. Both previous and updated theoretical value are shown in relation to the experimental value in Fig.~\ref{fig:SESEplot}.

\begin{table}
\begin{ruledtabular}
\begin{tabular}{llc}
Term & $g$-factor contribution & Ref. \\ 
\hline 
F-term & -4.083 5(1) & \cite{Sikora2020} \\ 
LAL & \phantom{-}0.108 6(35) & \cite{Sikora2020},TW \\ 
M-term, reducible & -0.059 9(11) & TW \\ 
M-term, N & \phantom{-}0.410 0(42) & TW \\ 
M-term, O & -0.206 5(169) & TW \\ 
P-term, reducible & -2.523 4 & TW \\ 
P-term, N \& O & \phantom{-}2.255 6(51) & TW \\ 
\hline
Sum & -4.099 2(185) & TW \\ 
\end{tabular} 
\end{ruledtabular}
\caption{Different contributions to the two-loop self-energy correction to the bound-electron $g$-factor in hydrogenlike tin according to Furry picture calculations. Contributions are given in units of $10^{-6}$. Where no uncertainty is given, all digits are significant.}
\label{tab:SESE}
\end{table}

\begin{table}
\begin{ruledtabular} 
\begin{tabular}{crlr}
Order & $g$-factor contribution && Ref. \\ 
\hline 
$(Z\alpha)^0$ & -3.713 89 & &\cite{Petermann1957,Sommerfield1958}\\ 

$(Z\alpha)^2$ & -0.082 40 & &\cite{Czarnecki00}\\ 

$(Z\alpha)^4$ & -0.658 41 & &\cite{Pachucki05} \\ 
 
$(Z\alpha)^5$ & \phantom{-}0.202 26& & \cite{Czarnecki2017} \\ 

$(Z\alpha)^{6+}$ & \phantom{-}0.000 00&(296 80) & \cite{Morgner2023} \\ 

\hline 
Sum & -4.252 44&(296 80) &\\ 

\end{tabular} 
\end{ruledtabular} 
\caption{Previous theory of the SESE correction to the bound-electron $g$-factor in H-like tin based on $(Z\alpha)$ expansion. Note that the $(Z\alpha)$ expansion converges very badly. Contributions expressed in units of $10^{-6}$.
}
\label{tab:SESENRQED}
\end{table}

\textit{Summary} -- We completed the calculation of the two-loop self-energy corrections to the bound-electron $g$-factor in hydrogenlike ions. The completed SESE theory allows us to significantly improve the overall theoretical accuracy of the bound-electron $g$-factor in the medium- to high-$Z$ regime. Our improved result also enables a more detailed test of virtual light-by-light scattering contributions~\cite{Morgner2023}. We demonstrated this by presenting an updated theoretical $g$-factor of hydrogenlike tin whose uncertainty is improved by almost one order-of-magnitude compared to the previous best theory and which is now significantly limited by nuclear effects. For even higher $Z>50$, uncertainties due to nuclear effects are expected to be significantly larger, while the uncertainty of the SESE effect can be expected to be comparable to this work. 
This paves the way to a $g$-factor theory which will be primarily limited by nuclear effects, or, in turn, can be applied to extract nuclear root-mean-square radii from experimental $g$-factors and to improved searches towards new physics effects~\cite{Debierre2020,Sailer2022}.

We thank N. Oreshkina, H. Cakir and J. Morgner for insightful discussions. Supported by the Deutsche Forschungsgemeinschaft (DFG, German Research Foundation) Project-ID 273811115 - SFB 1225.

\end{document}